# Enhancing the Plasmonic Hotspot Density via Structural Engineering of Multi-layered $MoO_3$-Ag-Au Systems Under Extreme Electronic Excitation Conditions for Ultra-Sensitive SERS Applications

Om Prakash[1], Sharmistha Dey[1], Mayur Khan[2], Abhijith T[3,4], Udai Bhan Singh[5*], Ambuj Tripathi[6], Santanu Ghosh[1*]

[1]*Nanostech Laboratory, Department of Physics, Indian Institute of Technology Delhi, New Delhi 110016, India.*

[2]*Department of Neutron and Ion Methods, Nuclear Physics Institute of Czech Academy of Sciences, Řež 130, Husinec- 25065, Czech Republic.*

[3]*Department of Physics, PSG Institute of Technology and Applied Research, Neelambur, Coimbatore, Tamil Nadu 641062, India.*

[4]*Department of Nanoscience and Technology, PSG Institute of Advanced Studies, Peelamedu, Coimbatore, Tamil Nadu-641004, India.*

[5]*Department of Physics, Deendayal Upadhyaya Gorakhpur University, Gorakhpur, 273009, India.*

[6]*Inter- University Accelerator Centre (IUAC), New Delhi 110016, India.*

*Email: santanu1@physics.iitd.ac.in , udaibhansingh123@gmail.com*

**Abstract**

Surface-enhanced Raman spectroscopy (SERS) is a potent, label-free method for highly sensitive molecular detection. We illustrate ion-beam engineering of $MoO_3$-Ag-Au multilayer plasmonic substrates to improve SERS performance. Orthorhombic α-$MoO_3$ microflakes were produced via chemical vapour deposition (CVD) on Si/$SiO_2$ substrates. Thin films of Ag (5 nm) and Au (5 nm) were thermally evaporated onto the $MoO_3$ flakes, and the samples were subjected to 100 MeV $Ag^{8+}$ swift heavy ion irradiation at fluences of $3 \times 10^{11}$ and $3 \times 10^{12}$ ions $cm^{-2}$. Irradiation causes dewetting of metal films, prompting structural and morphological changes that result in the formation of dispersed Ag-Au nanoparticles, enhanced surface roughness, and defect generation within the $MoO_3$ lattice. X-ray diffraction (XRD) verifies the α-$MoO_3$ phase; field emission scanning electron microscopy (FESEM) elucidates nanoparticle formation and surface reorganisation; Raman spectroscopy and X-ray photoelectron spectroscopy (XPS) disclose vibrational alterations and binding-energy shifts in Mo 3d, indicative of oxygen vacancies (V_O) and partial reduction of Mo. SERS measurements of molecular probes demonstrate significantly increased Raman intensities following ion irradiation. Finite-difference time-domain (FDTD) simulations assess localised surface plasmon resonance (LSPR) and near-field enhancement linked to the nanoparticle–flake configuration, while density functional theory (DFT) calculations of the electronic structure and density of states (DOS) validate the involvement of V_O in facilitating charge-transfer interactions. Experimental and theoretical evidence suggest that targeted swift-ion irradiation adjusts both electromagnetic and chemical enhancement mechanisms in $MoO_3$-Ag-Au multilayers, offering a reliable method for creating tunable, high-performance SERS substrates for ultrasensitive molecular detection.

# 1. Introduction

Surface-enhanced Raman scattering (SERS) is a non-invasive, label-free vibrational spectroscopy technique capable of detecting single molecules as well as complex molecular ensembles[1–3]. It offers exceptional spatial and temporal resolution, high sensitivity, and molecular-specific identification[4]. A key advantage of SERS lies in its ability to capture Raman vibrational signatures, distinct molecular fingerprints, enabling chemical identification with extraordinary sensitivity, down to the single-molecule level[5]. Amplification of the inherently weak Raman signals is typically achieved using materials that exploit two complementary mechanisms: electromagnetic (EM) enhancement and chemical (CE) enhancement[6–9]. EM enhancement arises from the excitation of localized surface plasmons, generating strong electromagnetic fields at the surfaces of plasmonic nanoparticles, which can increase Raman scattering cross sections by up to $10^{12}$-fold[10]. CE enhancement involves charge transfer between adsorbed molecules and the nanostructured surface, modulating molecular vibrations and amplifying Raman signals by up to $10^3$-fold[11]. These mechanisms can operate independently or synergistically, providing substantial Raman signal amplification[12,13]. Due to these capabilities, SERS has become a transformative tool across surface science, biomedicine, cellular analysis, and environmental monitoring, offering rapid, ultrasensitive molecular detection for fundamental and applied research[14,15].

Traditionally, SERS-based detection relies on electromagnetic enhancement achieved through the design and fabrication of tailored nanostructures composed of noble metals such as Ag, Au, and Cu, which has led to significant improvements in SERS performance[16–19]. Recently, semiconductors have emerged as promising alternative substrates, offering excellent spectral stability and biocompatibility[8,20,21]. However, their inherently low sensitivity has limited broader application. Studies have shown that introducing defects and controlling morphology can significantly enhance the Raman response of metal oxide SERS substrates[22].

Metal oxides are particularly attractive due to their physicochemical stability, high uniformity, tunable refractive index, and adjustable band gap, making them excellent candidates for SERS applications[23,24]. Among these, molybdenum trioxide ($MoO_3$) is a layered transition metal oxide, held together by weak van der Waals interactions, and exhibits multiple advantageous properties: variable oxidation states, chemical inertness, low cost, non-toxicity, and robust thermal and chemical stability[8,25]. The α-phase of $MoO_3$ is an n-type semiconductor with an orthorhombic crystal structure and a band gap of ~3.2 eV, representing the most thermodynamically stable polymorph under ambient conditions[26]. These characteristics make $MoO_3$ a promising platform for functional SERS substrates, where defect engineering, morphological tuning, and hybridization with plasmonic metals can be employed to overcome sensitivity limitations and achieve high-performance molecular detection[27,28].

Various techniques have been employed to fabricate SERS substrates, including self-assembly, nanoimprint lithography, electron-beam lithography, and nanosphere lithography[29–32]. However, these approaches are often complex, time-consuming, and costly, limiting their scalability for industrial applications[33]. An alternative and versatile method for tuning material properties is the introduction of engineered defects via high-energy heavy ion irradiation[34,35]. By carefully adjusting ion parameters such as type, energy, and fluence, it is possible to generate specific defect densities at controlled depths with high precision[36]. Ion irradiation and ion implantation have been successfully used to produce uniform SERS-active surfaces over large areas[37,38]. High-energy heavy ions traverse materials in nearly linear trajectories, penetrating several microns while affecting only a narrow cylindrical region of a few nanometers around their path, enabling precise surface and subsurface modifications. Recent studies have demonstrated the potential of this approach for tailoring nanostructures and enhancing SERS activity[34,39,40]. For example, Zhaoyi et al. reported 3 MeV $Cu^+$ irradiation of nanoporous copper ($3.36 \times 10^{14}$ ions $cm^{-2}$), resulting in increased SERS enhancement factors[41]. Similarly, Verma et al. observed morphological changes and reduced crystallinity in MoO$_3$ thin films following 100 MeV Ni$^{7+}$ irradiation at fluences ranging from $5 \times 10^{12}$ to $3 \times 10^{13}$ ions/cm$^2$, with Raman analysis confirming diminished peak intensity and spectral broadening[42]. Despite these advances, a systematic study of Ag ion irradiation on MoO$_3$ and MoO$_3$-Ag-Au hybrid multilayers for SERS applications remains unexplored, motivating the present work[43–45].

In this study, we investigate the effects of 100 MeV Ag$^{8+}$ swift heavy ion (SHI) irradiation on MoO$_3$ microflakes, MoO$_3$-Ag, and MoO$_3$-Ag-Au multilayer systems to elucidate the role of ion irradiation in modifying the morphology of MoO$_3$ and generating structural defects. To engineer plasmonic hotspots and induce electromagnetic enhancement, ultrathin Ag and Ag-Au films were deposited on the MoO$_3$ microflakes, which subsequently transformed into nanoparticle architectures upon SHI exposure. The structural and morphological evolution of these systems was characterized using field emission scanning electron microscopy (FESEM), grazing incidence X-ray diffraction (GIXRD), X-ray photoelectron spectroscopy (XPS), and Raman spectroscopy to understand the mechanisms underlying variations in SERS activity. The SERS performance of the fabricated substrates was evaluated using dye molecules, including Rhodamine 6G (R6G), Methyl Blue (MB), and Methyl Orange (MO), while thiram was detected at trace concentrations to demonstrate the substrate's applicability for pesticide sensing. Finite-difference time-domain (FDTD) simulations were conducted to model the plasmonic response, and density functional theory (DFT) calculations of the electronic structure and density of states (DOS) provided atomistic insight into ion-induced defects, predominantly oxygen vacancies, and their role in enhancing surface Raman activity. These combined experimental and theoretical analyses offer a comprehensive understanding of the interplay between ion irradiation, defect formation, and SERS enhancement in MoO$_3$-based multilayer systems.

## 2. Experimental Methods

### 2.1 Materials Used

Propanol ($C_3H_8O$, ≥99.9%), ethanol ($C_2H_5OH$, ≥99.8%), $MoO_3$ powder, Ag, Au, rhodamine 6G (R6G), methyl blue (MB), methyl orange (MO), thiram, and deionized (DI) water were obtained from commercial suppliers and used without further purification. For thin-film deposition, $SiO_2$ substrates with dimensions of 1 × 1 cm² were employed. Prior to deposition, the substrates were subjected to a standard cleaning protocol to ensure the removal of organic and particulate contaminants.

### 2.2. Fabrication of α-MoO₃ micro flakes and α-MoO₃ Ag-Au multilayer films

The $MoO_3$ microflakes were synthesized by the chemical vapour deposition (CVD) process on a $Si/SiO_2$ substrate using a microprocessor-programmable single-zone tubular furnace. The schematic diagram is illustrated in Figure 1a. $MoO_3$ powder (99.9% purity, particle size 30–50 nm, 21 mg) served as the precursor. The substrate was placed 13 centimetres away from the central zone of the furnace and kept at a temperature of 550 degrees Celsius. During the same time, the $MoO_3$ powder was heated to 850°C and kept in a quartz boat positioned within the central zone of the furnace. At a flow rate of 30 standard cubic centimetres per minute, argon was utilised as the carrier gas. To obtain the flakes, it is essential that the substrate be positioned at the proper distance from the central zone and at the appropriate elevation from the base. A 15 mm high boat was utilised to secure the substrate, as it will be positioned within the dense gas flow region. The furnace was heated at a rate of 12 °C per minute to reach 850 °C. The deposition lasted for 20 minutes, after which the heater was turned off for natural cooling. Subsequently, these samples were placed in a thermal evaporation system to deposit thin films of Ag and Ag-Au on $MoO_3$ microflakes. The chamber pressure was $3\times10^{-5}$ mbar during the deposition. An inbuilt thickness monitor in the system measured the thickness. 5nm Au and Ag films were deposited on the MoO$_3$ film. Pristine MoO$_3$, MoO$_3$-Ag, and MoO$_3$-Ag-Au samples were irradiated with 100 MeV Ag$^{8+}$ ions at varying fluences to investigate irradiation-induced morphological and chemical modifications as illustrated in Figure 1.

*Figure 1: Schematic representation of the growth of the MoO$_3$ microflakes, MoO$_3$-Ag, MoO$_3$-Ag-Au and subsequent ion-irradiation process and formation of nanostructures (NSs).*

## 2.3 Characterization of the Fabricated Substrate

The pristine and irradiated thin films were characterized using GIXRD (Phillips X'pert Pro) measurement with Cu Kα (λ = 1.54 Å) radiation in the range of 20–80° for the determination of crystal structure. Raman spectroscopy was performed using the Renishaw Micro Raman Spectroscope. A 532 nm wavelength diode laser was employed as the source. The chemical valence states of the samples were investigated by using X-ray photoelectron spectroscopy (XPS). XPS spectra were recorded using AXIS Supra with the monochromatic Al Kα X-ray source (1486.6 eV). The pass energy used was 20 eV, and the overall resolution was ∼ 0.6 eV. The pressure during the measurements was ∼3 × 10$^{-9}$ mbar. Furthermore, the morphology of different structures was studied using FESEM (TESCAN Magna LMU). The FDTD simulation (Ansys Lumerical) was performed to understand the plasmonic properties and field distribution over the substrate. To complement and interpret the experimental findings, the electronic structure and density of states (DOS) were investigated through first-principles calculations within the framework of DFT.

## 2.4 SERS Measurements

The SERS measurements were conducted employing the Renishaw Raman system with a 785 nm laser excitation. A solution of different dye molecules at various concentrations was drop-cast onto the SERS substrate and allowed to air dry prior to measurement. The laser power was calibrated to 20 mW to avert sample damage. A microscope objective (MO) with a magnification of 50× and a numerical aperture of 0.50. The spectral resolution was 1 cm$^{-1}$, with a spot diameter of approximately 0.2 µm and a depth of focus of about 5 µm. The spectra were acquired from a 1 × 1 µm² area with three accumulations and an exposure duration of 20 seconds. The baseline-corrected Raman signal was acquired from the system and subsequently utilised to compute the enhancement factor (EF) of the SERS substrate. The dye molecules, MO, MB, R6G, and pesticide Thiram molecules were utilized to evaluate the SERS performance.

## 3. Results and Discussion

### 3.1. Structural Evaluations

Figure 2 illustrates the XRD patterns obtained at ambient temperature for pristine and ion-irradiated samples. The XRD patterns confirm the formation of the pure α-MoO$_3$ phase [JCPDS 05–0508], characterised by an orthorhombic lattice structure with space group Pbnm. As discussed subsequently, the microstructural study of these samples clearly shows that MoO$_3$ flakes have different orientations on the substrate in pristine and ion-irradiated samples. Therefore, the diffraction peaks corresponding to the same (hkl) values in different samples do not have the same intensity as shown in Figure 2a. However, all samples crystallize in the pure α-MoO$_3$ phase, having an orthorhombic lattice structure. All peaks are identified with appropriate (hkl) values, which are denoted in Figure 2a. The XRD pattern of the MoO$_3$-Ag-Au multilayer is shown in Figure 2b, as illustrated, the Au and Ag films were very thin, so their subsequent peaks were not observed in the samples however α-MoO$_3$ peaks are clearly present and there is a decrease in the intensity because of ion-irradiation.

Figure 2c shows the Raman spectra of as-grown and annealed samples at room temperature. From the Raman spectrum, we can also conclude the formation of the pure α-MoO$_3$ phase. All peaks are identified with appropriate modes of vibration and denoted clearly in 2c, which arise due to the Mo-O bending and stretching vibrations, respectively. The peak at 995 cm$^{-1}$ corresponds to the stretching mode of terminal oxygen (Mo6$^+$=O), attributed to unshared oxygen. The peak at 817 cm$^{-1}$ corresponds to the stretching mode of doubly coordinated oxygen (Mo$_2$-O), resulting from corner-sharing oxygen atoms that are common to two octahedra. The peak at 664 cm$^{-1}$ corresponds to the triply coordinated oxygen (Mo$_3$-O) stretching mode, resulting from edge-sharing oxygen atoms prevalent among three octahedra. The peaks at 336 and 377 cm$^{-1}$ correspond to O-Mo-O bending modes. The peak at 283 cm$^{-1}$ corresponds to the wagging mode of the double bond Mo=O. The peaks at 197, 215, and 242 cm$^{-1}$ correspond to O=Mo=O twisting modes. The peaks at 157 and 291 cm$^{-1}$ correspond to the O=Mo=O wagging modes. The peaks at 469 cm$^{-1}$ correspond to the O-Mo-O stretching mode. The asterisk peak is recognised as the silicon substrate peak. The peak at 157 cm$^{-1}$ is attributed to the phonon wagging mode and the translation of the rigid chains. Also, it is to be noted that microflakes are oriented differently in various samples, which may influence the intensity of peaks observed in the X-ray diffraction patterns.

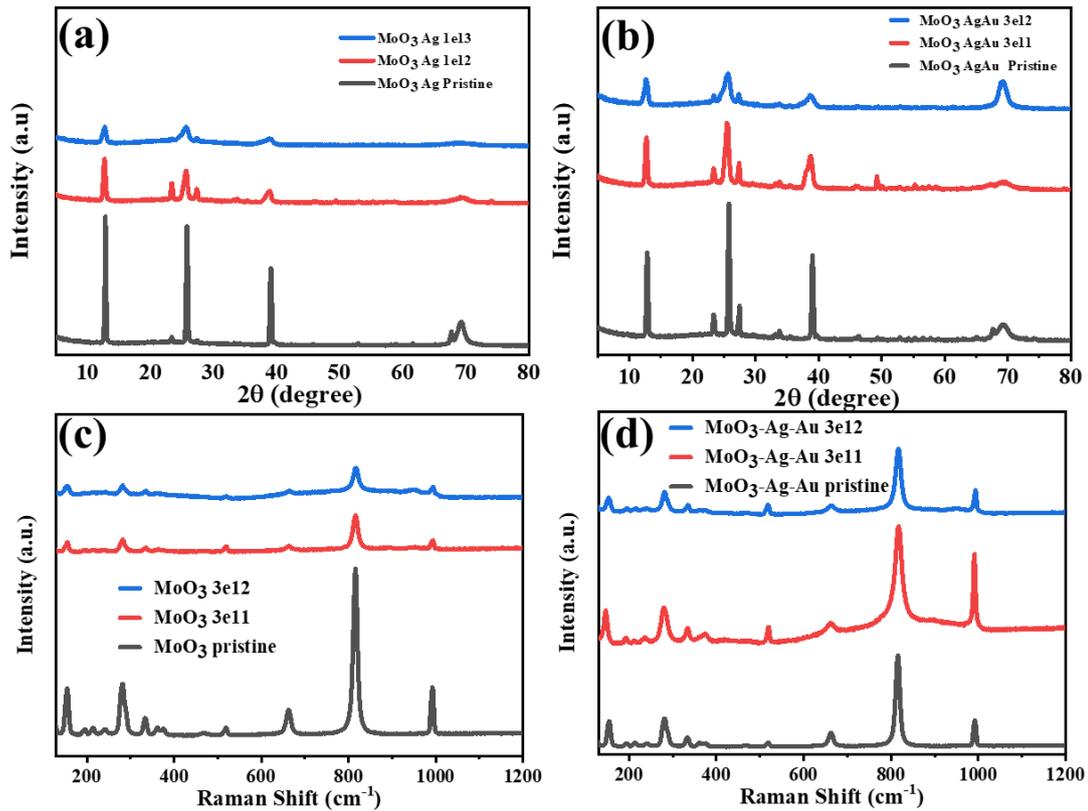

*Figure 2.* XRD patterns of (a) pristine and irradiated $MoO_3$ films, (b) pristine and irradiated $MoO_3$-Ag-Au films; Raman Spectra of (c) pristine and irradiated $MoO_3$ films, (d) pristine and irradiated $MoO_3$-Ag-Au films.

### 3.2 Morphological analysis (FESEM)

The morphology of the pristine and ion-irradiated substrates was examined by FESEM techniques. To understand the changes in surface structure and nanostructure size of different films, FESEM images are shown in Figure 3. For the $MoO_3$ micro flake, Figure 3a shows the morphology of the $MoO_3$ pristine developed by CVD, which confirms the deposition of micro flakes on top of the substrate. Morphological modifications occur due to the thermal spike caused by swift heavy ions, where intense localized electronic energy deposition results in transient melting and rapid quenching of the surface[39]. The change in the morphology was observed with a different fluence of ion irradiation, as shown in Figure 3b, c. It has been observed that because of ion-irradiation, the surface is modified, and cracks appear on it at higher fluences; also, melting of the microflakes takes place as shown in the insert images in Figure 3 b, c. The MoO$_3$-Ag-Au pristine sample FESEM image is shown in Figure 3d, which shows a smooth deposition of Ag-Au film on MoO$_3$ microflakes. Figure 3e, f corresponds to the MoO$_3$-Ag-Au substrate irradiated with 100 MeV Ag$^{8+}$ ions at fluences of $3 \times 10^{11}$ and $3 \times 10^{12}$ ions/cm$^2$, respectively. The insert image in Figure 3e is the higher magnification image corresponding to $3 \times 10^{11}$ ions/cm$^2$ fluence. This clearly shows the formation of Ag-Au nanostructure on the surface of the MoO$_3$ microflakes, leading to the formation of electromagnetic hotspots. The FESEM images corresponding to the MoO$_3$-Ag substrate are shown in Figure S1 of the supporting information, where nanostructure formation on the surface of the substrate is also observed due to ion irradiation.

**Figure 3.** *FESEM images of a) MoO$_3$ Pristine film, (b) MoO$_3$ $3 \times 10^{11}$ ions/cm$^2$ film, (c) MoO$_3$ $3 \times 10^{12}$ ions/cm$^2$ film, (d) MoO$_3$-Ag-Au Pristine film (e) MoO$_3$-Ag-Au $3 \times 10^{11}$ ions/cm$^2$ film, (f) MoO$_3$-Ag-Au $3 \times 10^{12}$ ions/cm$^2$ film.*

**3.3 X-ray photoelectron spectroscopy (XPS)**

To understand the reason behind the enhancement in the SERS signal with increasing irradiation fluences, the electronic structure of the samples has been studied. From FESEM images, the formation of nanostructures is the primary reason for the enhancement of the SERS signal. To understand the effect of vacancies, XPS data were elaborately studied. Oxygen vacancies in MoO$_3$ significantly improve SERS signals by creating defect states that promote effective charge transfer between the substrate and analyte molecules. These vacancies create additional active adsorption sites, leading to enhanced chemical interactions and increased Raman signal amplification[46,47]. In Figure 4, the Mo 3d core level XPS spectra is deconvoluted into 4 peaks with a pseudo-Voight function. The background signals are subtracted by using the Shirley background. To cancel the charging effect, the C 1s spectra at 284.6 eV is used as a reference. To fit the Mo 3d spectrum, the separation and the relative area ratio between the two spin-orbit doublets (5/2 and 3/2) are maintained as 3.2 eV and 1.5, respectively. As evident, $Mo^{6+}$ valence states predominate in all the samples, which confirms pure $MoO_3$ phase formation. $Mo^{6+}$ $3d_{5/2}$ and $3d_{3/2}$ are located at 232.6 and 235.8 eV, and lower valence states of Mo ($Mo^{5+}$) $3d_{5/2}$ and $3d_{3/2}$ are located at 232 and 235.2 eV, respectively. The presence of $Mo^{5+}$ states confirms the formation of vacancies after irradiation. From Table 1, it is confirmed that the $MoO_3$-Ag-Au irradiated at $3 \times 10^{11}$ ions/cm² sample shows the highest content of $Mo^{5+}$, and it shows the highest SERS signal; there is a one-to-one correspondence between vacancies and the enhancement in the SERS signal.

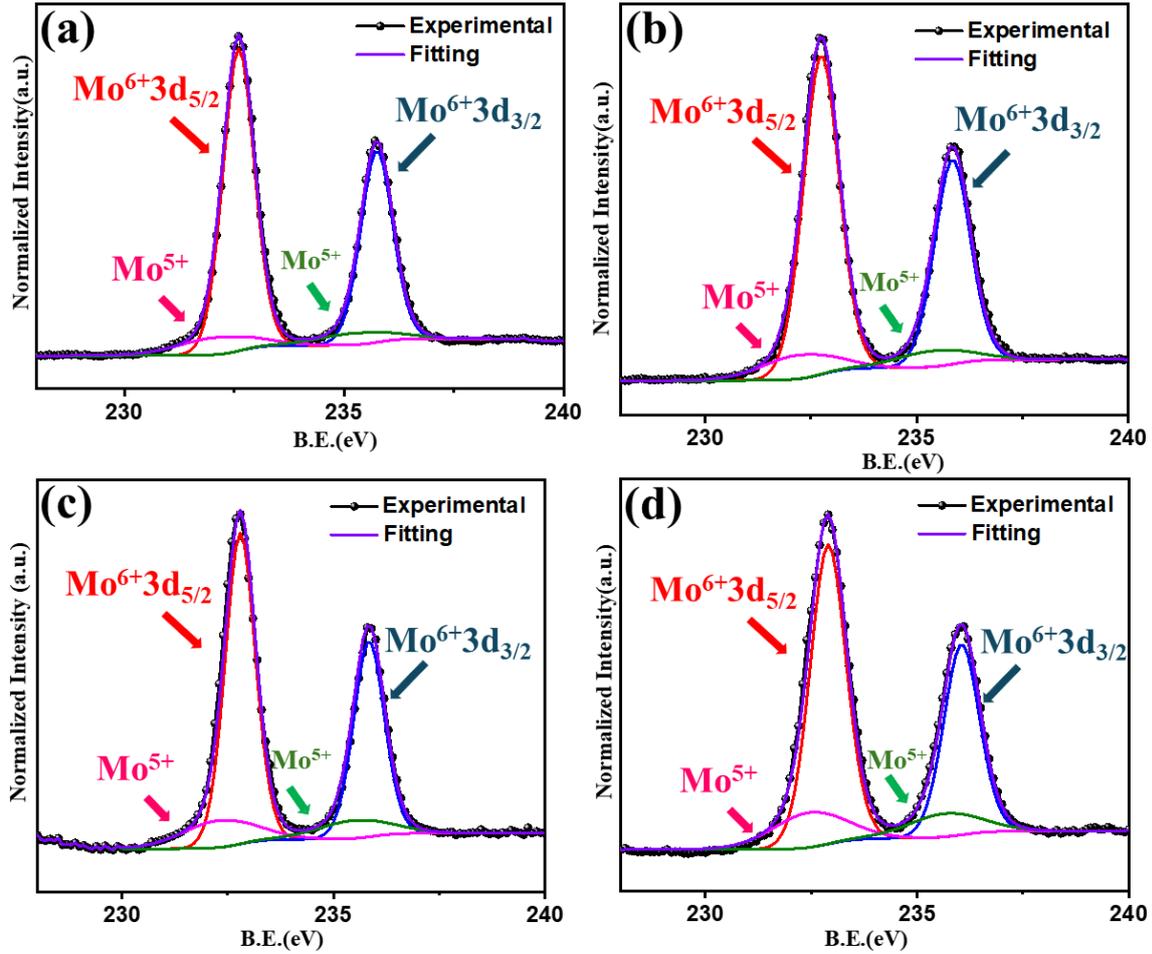

*Figure 4.* Mo 3d core level XPS spectra of (a) $MoO_3$ Pristine film, (b) $MoO_3$ $3 \times 10^{11}$ ions/cm² film, (c) $MoO_3$ Ag-Au-Pristine film, and (d) $MoO_3$-Ag-Au at $3 \times 10^{11}$ ions/cm² film samples. $Mo^{6+}$ states are predominant in all the samples. Peaks associated with $Mo^{5+}$ states indicate the existence of oxygen vacancies in all samples.

Table 1. Mo Content in different oxidation states for the samples

| Samples | $Mo^{6+}$ content (%) | $Mo^{5+}$ content (%) |
|---|---|---|
| $MoO_3$-Pristine | 88.06 | 11.93 |
| $MoO_3$-$3\times10^{11}$ | 86.89 | 13.11 |
| $MoO_3$-Ag-Au-Pristine | 82.73 | 17.27 |
| $MoO_3$-Ag-Au-$3\times10^{11}$ | 80.83 | 19.17 |

## 3.4 Optimization and Versatility of SERS Substrate

The Raman intensities were analysed to determine optimal performance. The $MoO_3$, $MoO_3$-Ag, and $MoO_3$-Ag-Au substrates were synthesized and irradiated with 100 MeV $Ag^{8+}$ ions at fluences of $3 \times 10^{11}$ and $3 \times 10^{12}$ ions/cm² to investigate their SERS performance using methylene blue (MB) as a model analyte. Remarkably, the $MoO_3$-Ag-Au substrate irradiated at $3 \times 10^{11}$ ions/cm² exhibited the highest Raman intensity and a detection limit as low as $10^{-12}$ M for MB. This outstanding enhancement arises from the synergistic interplay between ion-induced oxygen vacancies in $MoO_3$, which promote efficient charge transfer, and the strong electromagnetic field generated by the closely spaced Ag and Au nanoparticles. At this optimal fluence, the nanostructure remains well-preserved, ensuring a high density of SERS-active "hot spots" and efficient analyte adsorption. In contrast, higher fluence led to excessive defects and structural damage, diminishing SERS activity. SERS signals were detected for MB at the characteristic peaks of 451 cm$^{-1}$ (C–S–C bending), 601 cm$^{-1}$ (C-S stretching, ring Deformation), 774 cm$^{-1}$ (C–H out-of-plane bending), 1184 cm$^{-1}$ (C–N stretching, aromatic bend), 1305 cm$^{-1}$ (C–N stretching, aromatic ring), 1398 cm$^{-1}$ (Aromatic ring, C-N stretch), 1626 cm$^{1}$ (C=C stretching). These results demonstrate that precise ion beam engineering is crucial for maximizing SERS sensitivity in multi-component hybrid substrates.

*Figure 5. Raman spectra of the MB molecule on $MoO_3$, $MoO_3$-Ag, and $MoO_3$-Ag-Au substrates, Pristine and irradiated with 100 MeV $Ag^{8+}$ ions at fluences of $3 \times 10^{11}$ and $3 \times 10^{12}$ ions/cm².*

To further confirm the hotspot presence on the substrate the *E*-field intensity enhancement profiles on pristine and irradiated $MoO_3$-Ag-Au films were monitored using theoretical simulations. The simulations were carried out using Ansys Lumerical FDTD software. The simulation structures were modeled using the methods as reported in our earlier study[7]. In this work, the $MoO_3$-Ag-Au structures were modelled using the experimentally obtained parameters. The *E*-field intensity enhancement profiles for pristine and irradiated $MoO_3$-Ag-Au films are shown in Figure 6a and 6b, respectively. There was no significant enhancement in field intensity observed in the case of the pristine $MoO_3$-Ag-Au structure. On the other hand, the field intensity significantly enhanced throughout the surface of the irradiated $MoO_3$-Ag-Au surface. The dark red spots indicate the localized plasmonic hotspots, which are beneficial to carry a large amount of analyte molecules and enhance their Raman scattering cross-section. To understand the extent of the E-field intensity enhancement factor ($|E|^2/|E0|^2$) on both surfaces, the E-field intensity line profiles (profiles 1 and 2) were recorded from 2D profiles. Plots for the variation of $|E|^2/|E_0|^2$ with respect to the distance are shown in Figure 6c. Compared to the pristine substrate, the irradiated substrate showed plenty of plasmonic hotspots with high field intensity enhancement, indicating the viability of the present SERS substrates in molecular sensing applications.

*Figure 6. The E-field intensity enhancement 2D profiles at the top surface of (a) pristine $MoO_3$-Ag-Au film, and (b) $MoO_3$-Ag-Au film irradiated with 100 MeV $Ag^{8+}$ ions at fluences of 3 × $10^{11}$ ions/$cm^2$ (c) The E-field-intensity line profiles recorded from the respective 2D E-field intensity enhancement profiles. The dashed lines in (a) & (b) indicate the location at which the field intensity is measured.*

To assess the versatility of the above SERS substrate, additional dye molecules were detected, and the pesticide Thiram was also identified on the optimized SERS substrate at a very low concentration. All the molecules were detected by the optimised SERS substrate at a concentration of 0.1 μM. For this, $MoO_3$-Ag-Au irradiated with 100 MeV $Ag^{8+}$ ions at fluences of $3 \times 10^{11}$ ions/cm² was employed as a highly effective substrate for SERS detection of R6G, MO, and Thiram molecules, as shown in Figure 7a. The formation of Ag-Au nanoparticles on $MoO_3$ flakes via high-energy ion irradiation results in a highly effective SERS substrate that leverages both electromagnetic and chemical enhancement mechanisms. The irradiation process induces the growth of uniformly distributed bimetallic nanoparticles, which generate intense and consistent electromagnetic "hotspots" due to localized surface plasmon resonance, thereby significantly amplifying Raman signals. Simultaneously, the ion irradiation introduces oxygen vacancies within the $MoO_3$ lattice, as confirmed by spectroscopic analyses, which facilitate charge transfer interactions between the substrate and adsorbed analyte molecules, further boosting the SERS response through chemical enhancement. This synergistic combination of uniform plasmonic nanostructures and engineered defect sites yields a substrate with exceptional sensitivity, uniformity, and stability, making it highly advantageous for ultrasensitive molecular detection applications. Strong SERS signals were detected for R6G at the characteristic peaks of aromatic C-C-C ring in plane vibration (611 $cm^{-1}$), C-H out of plane vibration (775 $cm^{-1}$), N-H in-plane vibration (1308 $cm^{-1}$), C–C and aromatic stretching vibrations (1361 $cm^{-1}$, 1505 $cm^{-1}$, and 1651 $cm^{-1}$), as illustrated in Figure 7a. It was discovered that R6G's enhancement factor (EF) on these substrates ranged from $10^7$ to $10^8$, allowing for detection limits as low as 0.1 pM. There were distinct SERS peaks observed in methyl orange at 701 $cm^{-1}$ (representing C–N stretching), 1181 $cm^{-1}$ (representing N=N stretching), 1377 $cm^{-1}$ (representing C–H bending), and 1601,1625 $cm^{-1}$ (representing aromatic ring vibrations), as shown in Figure 7a. Thiram, a sulfur-containing pesticide, was identified by its distinctive peaks at 583 cm$^{-1}$ (S–S stretching), 625 cm$^{-1}$ (C–S stretching), 657 cm$^{-1}$ and 704 cm$^{-1}$ (C–H deformation), 814 cm$^{-1}$ and 1211 cm$^{-1}$ (N-CH3 stretching), and 1592 cm$^{-1}$ and 1616 cm$^{-1}$ (CH$_3$ rocking, C-N stretching), as illustrated in Figure 7a. Figure 7b demonstrates unequivocally that the SERS substrate that was developed is capable of detecting analyte molecules at extremely low concentrations, with the successful identification of Thiram molecules at concentrations as low as 0.1 pM. It is believed that the synergistic effect of uniform plasmonic Ag-Au nanoparticle hotspots and the presence of oxygen vacancies in the MoO$_3$ flakes are responsible for this remarkable sensitivity of MoO$_3$-Ag-Au, irradiated with 100 MeV Ag$^{8+}$ ions at a fluence of 3 × 10$^{11}$ ions/cm$^2$. The fact that even minute quantities of the analyte can be reliably detected is an example of the substrate's extraordinary potential for applications, which includes ultra-sensitive molecular sensing. Additionally, the substrate functions as a versatile and ultrasensitive platform for the multiplex detection of dye and pesticide molecules, which offers a significant amount of potential for applications in biomedical diagnostics, food safety, and environmental monitoring.

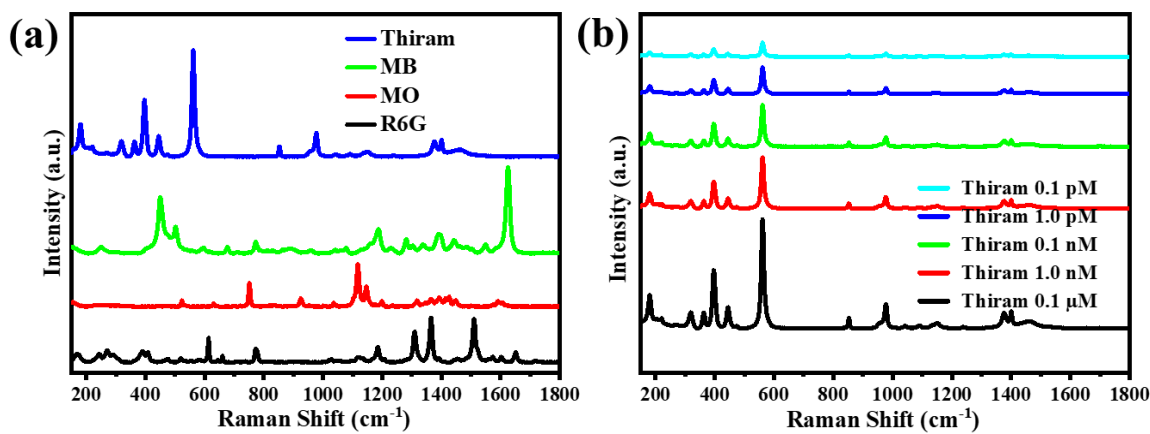

*Figure 7.* Raman spectra of the (a) R6G, MO, MB, and Thiram molecules on MoO$_3$-Ag-Au irradiated with 100 MeV Ag$^{8+}$ ions at fluences of 3 × 10$^{11}$ ions/cm$^2$ at 0.1 μM concentration, (b) Detection of Thiram at different concentrations.

Furthermore, to verify the uniformity of the SERS signal across a 3 × 3 µm² area within a random location, Raman mapping was performed on the optimised SERS substrate MoO$_3$-Ag-Au irradiated with 100 MeV Ag$^{8+}$ ions at fluences of 3 × 10$^{11}$ ions/cm². This was done to ensure that the signal was consistent across the entire area. The duration of the exposure was reduced to ten seconds to minimize the total time required for signal collection. Additionally, the laser power was adjusted to thirty milliwatts. Figure 7a-c depicts the Raman mapping of three distinct peaks of the R6G molecule, which are 1505, 1361, and 611 cm$^{-1}$. This illustration demonstrates that sufficient uniformity has been achieved. As a result of the presence of hotspots, which result in the confinement of electric fields, a significant SERS signal was obtained. The plot of Raman spectra obtained from all 169 pixels is depicted in Figure 7d. This plot demonstrates that the substrate is sufficiently uniform. To determine whether the SERS substrate is reproducible, the Raman peaks of R6G at 611 cm$^{-1}$, 1361 cm$^{-1}$, and 1505 cm$^{-1}$ were considered. The Raman intensities at random hot spots had standard deviations of 13%, 12%, and 15%, respectively, for peaks at 611 cm$^{-1}$, 1361 cm$^{-1}$, and 1505 cm$^{-1}$. These values were determined by standard deviations. It is possible to consider the SERS substrate to be extremely uniform.

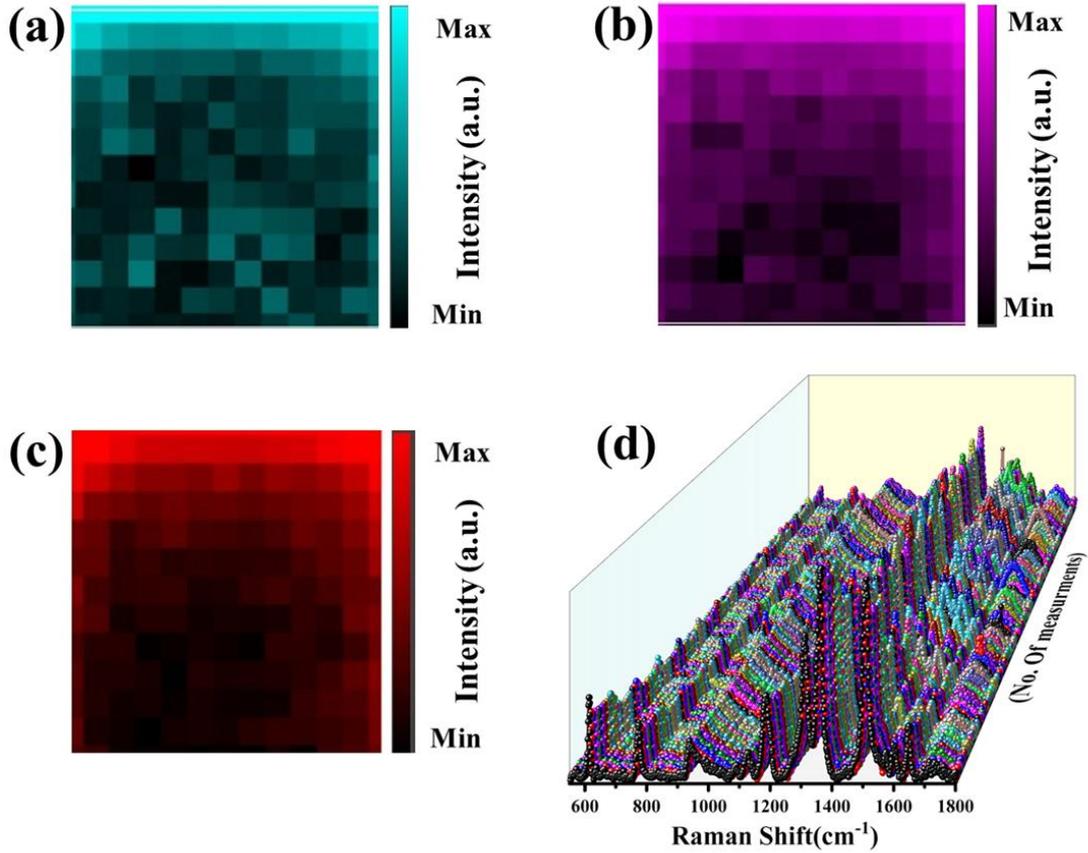

*Figure 8.* 2-D Raman mapping images of the $10^{-9}$ M R6G molecules over the substrate for (a)1505 cm$^{-1}$, (b) 1361 cm$^{-1}$, and (c) 611 cm$^{-1}$ Raman modes, respectively. (d) Raman spectra of each point in the measurement region.

**3.5 Enhancement Factor Calculation**

The enhancement factor (EF) is determined by calculating the results obtained from the 0.1nM Thiram solution on both the optimised SERS substrate and the bare silicon substrate, as depicted in Figure S2. Through the utilisation of the following formula, the enhancement factor was computed.

$$EF = \frac{I_{SERS} \times N_{RAMAN}}{I_{RAMAN} \times N_{SERS}} \quad (1)$$

where $I_{Raman}$ and $I_{SERS}$ represent the intensities of the vibrational mode of Thiram that are absorbed on bare silicon and SERS substrate, respectively; $N_{Raman}$ and $N_{SERS}$ represent the number of the probe molecules that are absorbed on the bare silicon surface and SERS surface, respectively.

$$N_{Raman} = \frac{Ah\rho N_A}{M} \quad (2)$$

$$N_{SERS} = \frac{cVAN_A}{s} \quad (3)$$

where A represents the area of the incident laser spot on the substrate surface, h denotes the depth of laser penetration, $\rho$ is the density of the probe molecule, and M indicates the molecular weight. $N_A$ represent the Avogadro number. V, c, and s denote the volume, molar concentration, and surface area of the Thiram solution, respectively. The computed EF for distinct Raman peaks of probe molecules with their modes of vibrations are recorded in Table 2. The experimental parameters remained constant for all the measurements[11].

*Table 2. Assigned Raman modes of the R6G molecule and the enhancement factor corresponding to MoO$_3$-Ag-Au irradiated with 100 MeV Ag$^{8+}$ ions at fluences of 3 × 10$^{11}$ ions/cm$^2$ for 10$^{-9}$M concentration.*

| Sr. No. | Raman Peak (cm$^{-1}$) | Modes of vibration | Enhancement factors (EFs) |
|---|---|---|---|
| 1 | 611 | Aromatic C-C-C ring in-plane vibration | 6.68 × 10$^7$ |
| 2 | 775 | Out-of-plane C-H vibration | 5.83 × 10$^7$ |
| 3 | 1036 | In-plane C-H vibration | 4.43 × 10$^7$ |
| 4 | 1185 | In-plane bending of C-H | 1.42 × 10$^8$ |
| 5 | 1361 | C-C Aromatic stretching | 2.43 × 10$^8$ |
| 6 | 1505 | C-C Aromatic stretching | 1.53 × 10$^8$ |
| 7 | 1651 | C-C Aromatic stretching | 3.42 × 10$^7$ |

## 4. DFT calculations

The electronic structure and density of states (DOS) were examined through first-principles calculations using Density Functional Theory (DFT) to complement and understand the experimental results. These simulations not only corroborate the experimental observations for ion-induced defects, mainly O-atom vacancies, but also provide an atomistic-level understanding of the material's surface-enhanced characteristics.

### 4.1 Structural details of α-MoO$_3$

The orthorhombic system with space group Pnma (No. 62) is where α-MoO$_3$ crystallizes. Approximately a ≈ 3.96 Å, b ≈ 13.86 Å, and c = 3.70 Å are the unit cell parameters[48]. Six O atoms with varying bond lengths coordinate each Mo atom in its deformed MoO$_6$ octahedra structure. Weak van der Waals interactions hold these octahedra together as they create bilayer sheets stacked along the [010] direction by sharing edges and corners. α-MoO$_3$ is relevant for electronic, catalytic, surface enhancing and sensing applications because of its layered organization, which produces noticeable structural anisotropy and enables intercalation and surface reactivity[49].

### 4.2 Computational details

All first-principles calculations were carried out within the framework of density functional theory (DFT) using the Quantum Espresso package[50–53]. The ionic cores were described with the projector augmented-wave (PAW) method, and the Perdew-Burke-Ernzerhof (PBE) form of the generalized gradient approximation (GGA) was employed for exchange–correlation effects[54,55]. A plane-wave cutoff of 65 Ry was used, with the electronic self-consistency threshold set to $10^{-7}$ eV. Structural relaxations were performed until the residual atomic forces were below $10^{-4}$ Ry/Bohr and the residual stress was below $10^{-4}$ kbar. For the pristine orthorhombic α-MoO$_3$ (space group *Pnma*), both lattice parameters and internal coordinates were relaxed. A Γ-centered 8×3×8 k-point mesh was used for geometry optimization and total-energy calculations, while a denser 16×6×16 mesh was applied for density of states (DOS) to capture fine features near the Fermi level. Band structures were calculated along the standard high-symmetry path of the orthorhombic Brillouin zone, and smooth dispersions were obtained by interpolation of eigenvalues between the k-points.

Oxygen vacancies were introduced using a supercell approach, where the primitive cell was expanded, and one O atom was removed to generate a neutral vacancy. The defective structures were fully relaxed with the same convergence criteria as above. For these supercells, a Γ-centred 3×2×3 k-point grid was adopted for relaxation, while denser meshes were used for DOS and PDOS. The electronic DOS and projected/partial DOS were calculated with Gaussian smearing, with orbital contributions resolved by projecting the plane-wave states onto PAW spheres (reporting primarily Mo-d and O-p states).

**4.3 Electronic Structure of Pristine and Oxygen-Deficient MoO$_3$**

To understand how oxygen vacancies influence the optical response and Raman enhancement behaviour of MoO$_3$, we carried out calculations of the band structure, total density of states (DOS), and partial density of states (PDOS) for both the pristine and oxygen-deficient systems as shown in Figure 9a-f. These results clearly highlight how vacancy-driven modifications in the electronic structure are closely connected to the observed surface-enhanced Raman scattering (SERS) activity. The computation of the electronic structure and density of system states also relies on the lattice relaxation. By choosing the system's minimal energy, the relaxed lattice parameters for each system have been optimized[56–58]. This was achieved by running the computation with different experimental lattice parameter values on both the increasing and decreasing sides[57,58]. The optimized lattice parameters for MoO$_3$ were a = 3.91 Å, b = 14.22 Å, and c = 3.68 Å are the unit cell parameter, respectively. These values are reasonably similar to the experimental values as demonstrated by G. Chenjie[22] et al. and as determined in previous studies[59].

For pristine α-MoO₃, the calculated band structure shows an indirect semiconducting gap of about 1.90 eV, as shown in Figure 9a, which agrees well with earlier theoretical and experimental reports[59]. The DOS and PDOS analyses indicate that the valence band maximum is dominated by O-2p orbitals, while the conduction band minimum originates mainly from Mo-4d states, as shown in Figures 9b and 9c, respectively. This distribution reflects strong Mo-O hybridization and confirms the charge-transfer nature of the electronic transitions. Introducing an oxygen vacancy leads to notable changes in the electronic structure. Defect-induced mid-gap states appear in the band structure, which are primarily localized Mo-4d states associated with reduced $Mo^{5+}$ centers. The DOS and PDOS results further verify the rise in electronic density near the Fermi level from Figures 9e and 9f, respectively. Consequently, the effective band gap narrows from 1.90 eV (pristine) to around 1.09 eV, creating additional absorption pathways in the visible spectral range.

### 4.4 Correlation with Ion Beam Irradiation

Ion-beam irradiation offers a precise way to generate and control oxygen vacancies in $MoO_3$. By adjusting parameters such as ion energy and irradiation dose, the vacancy concentration can be systematically tuned. Our simulations (band structures, DOS and PDOS) indicate that these vacancies introduce defect states which enhance both electrical conductivity and visible-light absorption. Experimentally, irradiation has also been shown to alter the oxidation state of Mo and increase the number of reduced sites, thereby modifying the electronic response of the material. The observed link between irradiation impact and defect-mediated electronic transitions is consistent with our DOS and PDOS calculations.

### 4.5 Implications for Surface-Enhanced Raman Scattering (SERS)

The mid-gap states created by oxygen vacancies play a crucial role in boosting the SERS response. In pristine MoO$_3$, the wide band gap restricts charge-transfer (CT) processes between the substrate and adsorbed molecules under commonly used laser excitations. With the introduction of vacancies, intermediate energy levels appear, providing resonant pathways for photon-driven electron transfer. These states bridge the energy mismatch and facilitate stronger Raman signal enhancement. Thus, ion-beam irradiation establishes a direct connection between defect formation and SERS activity. At low to moderate fluences, vacancy generation enhances CT and improves sensitivity. However, beyond an optimal defect concentration, clustering of vacancies can lead to non-radiative recombination and diminish stability of the Raman signal. This results in a volcano-like trend, where SERS enhancement increases up to a certain point and then declines at higher irradiation doses. This is also consistent with the exiting studies.[46] Overall, the integration of band-structure analysis with DOS/PDOS calculations highlights oxygen vacancies as the primary lever for tuning MoO$_3$'s electronic properties toward SERS applications. Ion-beam methods, by enabling precise defect engineering, provide robust approach to optimize charge-transfer efficiency and design advanced semiconductor substrates for molecular sensing.

***Figure 9.*** *MoO$_3$ pristine (a) Band structure (b) Density of states (DOS) (c) Partial density of states (PDOS) and MoO$_3$ with O-vacancy (d) Band structure (e) Density of states (DOS) (f) Partial density of states (PDOS).*

## 5. Conclusion

In summary, we have reported the synthesis of MoO$_3$ microflakes, MoO$_3$-Ag, and MoO$_3$-Ag-Au multilayer systems as SERS substrates. This system was employed to swift heavy ion-irradiation with 100 MeV Ag$^{8+}$ ions at fluences of $3 \times 10^{11}$ and $3 \times 10^{12}$ ions/cm$^2$. It was found that these substrates can provide highly sensitive detection of distinct dye molecules and pesticides at minute concentrations. The FESEM confirms that SHI ion-irradiation induced the defects and oxygen vacancies, and the Ag-Au nanostructure is formed on top of the $MoO_3$ microflakes. The XPS study reveals the presence of $Mo^{5+}$ states, confirming the formation of vacancies in the lattice after irradiation, which contributes to oxygen vacancies. The electric field distribution obtained using FDTD reveals the uniformity of the electric field distribution on top of the SERS substrate, as well as the presence of hotspots. The electronic structure and density of states (DOS) were investigated using first-principles calculations in Density Functional Theory (DFT). These simulations confirm experimental observations of ion-induced defects, mostly O-atom vacancies, and provide an atomistic-level understanding of the material's surface-enhanced properties. SERS substrate uniformity was also confirmed with Raman mapping corresponding to R6G molecules for 1505, 1361, and 611 $cm^{-1}$ characteristic peaks. The synergistic effect of electromagnetic and chemical enhancement leads to the $MoO_3$-Ag-Au sample irradiated with 100 MeV $Ag^{8+}$ ions at fluences of $3 \times 10^{11}$ ions/$cm^2$, giving the maximum enhancement with an enhancement factor up to $2.43 \times 10^8$. Also, the distinct molecules MO, MB, and Thiram were successfully detected at very low concentrations up to 0.1 pM. These results confirm the versatility of the SERS substrate for wide area uniformity and demonstrate considerable potential as a cost-effective, dependable, and portable sensing platform for applications in environmental monitoring, food safety, and biomedical diagnostics.

**Supporting Information**

FESEM images of $MoO_3$-Ag thin film; Enhancement factor calculation.

**Acknowledgments**

The author, Mr. Om Prakash, would like to thank UGC for funding the fellowship. One of the authors, Udai Bhan Singh, is thankful to ANRF for financial support through Project No. SUR/2022/004428. We express our gratitude to the Inter-University Accelerator Centre (IUAC) in New Delhi for the high-energy ion beam facility. Central Research Facility and Nanoscale Research Facility at IIT Delhi for their characterisation facilities.

Notes

The authors declare no competing financial interest.

Author Information

Corresponding Authors

**Santanu Ghosh** - *Nanostech* Lab*, Department of Physics, Indian Institute of Technology Delhi, New Delhi 110016, India*
Email: santanu1@physics.iitd.ac.in

**Udai B. Singh**- *Department of Physics, Deendayal Upadhyay Gorakhpur University, Gorakhpur, 273009, India.*

Email: udaibhansingh123@gmail.com

Author Contributions

**Om Prakash**: Conceptualization, Investigation, Formal analysis, Simulations, Data curation, Writing original draft. **Sharmistha Day**: Investigation, Formal analysis, Data curation, Writing original draft. **Mayur Khan:** Formal analysis, Simulations, Data curation, Writing original draft. **Abhijith T:** Formal analysis, Simulations, Data curation, Writing original draft. **Udai Bhan Singh:** Supervision, Conceptualization, Investigation, Writing original draft. **Ambuj Tripathi:** Investigation, Data curation, Writing original draft. **Santanu Ghosh:** Supervision, Conceptualization, Investigation, Simulations, Data curation, Writing original draft